\documentclass{PoS}

\usepackage{bm}
\usepackage{amsmath}

\newcommand{\tr}{\mathrm{tr}}
\newcommand{\I}{\mathrm{i}}
\def\Eq#1{Eq.~(\ref{#1})}
\def\Fig#1{Fig.~\ref{#1}}

\let\OLDthebibliography\thebibliography
\renewcommand\thebibliography[1]{
  \OLDthebibliography{#1}
  \setlength{\parskip}{0pt}
}

\title{Strange nucleon form factors and isoscalar charges with $N_f=2+1$ $\mathcal{O}(a)$-improved Wilson fermions}

\ShortTitle{Strange nucleon form factors and isoscalar charges}

\author{Dalibor Djukanovic$^{2}$, Harvey Meyer$^{1,2}$, Konstantin Ottnad$^{1,2}$, Georg von Hippel$^1$, \speaker{Jonas Wilhelm}$^1$, Hartmut Wittig$^{1,2}$\\
		\llap{$^1$}PRISMA$^+$ Cluster of Excellence and Institute for Nuclear Physics\\        
        Johannes Gutenberg University of Mainz\\
		Johann-Joachim-Becher-Weg 45\\
		55128 Mainz\\
		\llap{$^2$}Helmholtz Institute Mainz\\
		Staudingerweg 18\\
		55128 Mainz\\
		\newline
        E-mail: {\email{D.Djukanovic@him.uni-mainz.de}, \email{meyerh@kph.uni-mainz.de}, \email{kottnad@uni-mainz.de}, \email{hippel@uni-mainz.de}, \email{jonas.wilhelm@uni-mainz.de}}, \email{wittig@kph.uni-mainz.de}}

\abstract{We report on our calculation of the strange contribution to the vector and axial vector form factors. The strange charge radii, magnetic moment, and axial charge are extracted by model independent $z$-expansion fits to the $Q^2$-dependence of the respective form factors. Furthermore, the isoscalar contribution to the axial and tensor charge is investigated by combining the calculation of connected and disconnected diagrams. The required renormalization is performed with the Rome-Southampton method. We make use of the CLS $N_f=2+1$ $\mathcal{O}(a)$-improved Wilson fermion ensembles. Results are reported for pion masses in the range $m_\pi=200-360\,\text{MeV}$ and lattice spacings $a=0.05-0.086\,\text{fm}$.}

\FullConference{37th International Symposium on Lattice Field Theory - Lattice2019\\
		16-22 June 2019\\
		Wuhan, China}

\begin{document}

\section{Introduction}
Recent experimental efforts have been devoted to measurements of the weak mixing angle in electron-proton scattering \cite{Androic,Becker}. Precise measurements at low energy are of high interest due to the sensitivity of the running of the weak mixing angle in extensions of the Standard Model \cite{Davoudiasl}. In order to control the effects of proton structure in the scattering experiments, knowledge of nucleon form factors is required. Among others, the strange form factors $G_E^s(Q^2)$, $G_M^s(Q^2)$ and $G_A^s(Q^2)$ need to be known precisely, which is a challenging task in experiments \cite{Pate}. Here, we extract these form factors from a first-principles Lattice QCD study, based on our previous work \cite{Djukanovic}.  Furthermore, we calculate the contributions of the $u$-, $d$- and $s$-quarks to the axial charge. These correspond to the contributions of the intrinsic spin of the respective quarks to the total spin of the nucleon \cite{Ji}. Following the same procedure, we also obtain the $u$-, $d$- and $s$-contributions to the tensor charge of the nucleon. 

\section{Extracting Form Factors from Lattice QCD}
To extract nucleon form factors, we consider the following ratio of nucleon three- and two-point functions \cite{Brandt}
\begin{align}
\begin{split}
R_{J_\mu}(\vec{q};\vec{p}^\prime;\Gamma_\nu) &= \frac{C_{3,J_\mu}^N(\vec{q},z_0;\vec{p}^\prime,y_0;\Gamma_\nu)}{C_2^N(\vec{p}^\prime,y_0)}\sqrt{\frac{C_2^N(\vec{p}^\prime,y_0)C_2^N(\vec{p}^\prime,z_0)C_2^N(\vec{p}^\prime\text{-}\vec{q},y_0\text{-}z_0)}{C_2^N(\vec{p}^\prime\text{-}\vec{q},y_0)C_2^N(\vec{p}^\prime\text{-}\vec{q},z_0)C_2^N(\vec{p}^\prime,y_0\text{-}z_0)}}\\[0.25cm]	
	&\stackrel{\text{s.d.}}{=} M_{\nu\mu}^1(\vec{q},\vec{p}^\prime) G_1(Q^2) + M_{\nu\mu}^2(\vec{q},\vec{p}^\prime) G_2(Q^2)\ ,
\end{split}
\label{eq:ratio}
\end{align}
where the second line corresponds to the spectral decomposition at large Euclidean times. At each value of $Q^2$, the ratios belonging to different $\mu$, $\nu$ and momenta $\vec{q}$, $\vec{p}^\prime$ can be grouped into a vector $\bm{R}$. Ordering the kinematic factors $M^1$ and $M^2$ accordingly into a matrix $M$, a (generally overdetermined) system of equations can be defined
\begin{equation}
M\ \bm{G} = \bm{R}\ \ ;\ \ M = \left( \begin{array}{c}
M^1_1\\
\vdots\\
M^1_N\\
\end{array}\ \begin{array}{c}
M^2_1\\
\vdots\\
M^2_N\\
\end{array} \right)\ \ ,\ \ \bm{G} = 
\left(
\begin{array}{c}
G_1\\
G_2\\
\end{array}
\right),\ \ \bm{R} = 
\left(
\begin{array}{c}
R_1\\
\vdots\\
R_N\\
\end{array}
\right)\ .
\label{eq:system}
\end{equation}

In the case of the disconnected contributions, we reduce the size of the system by first averaging equivalent three-point functions, where the momenta at the source and the sink are related by spatial symmetry \cite{Syritsyn}, followed by a construction of the ratios from the resulting averages. Concerning the connected contributions, we perform an average of equivalent ratios. In both cases we drop non-contributing ratios, for which both kinematic factors $M^1$ and $M^2$ are zero. Furthermore, we average the two-point functions over equivalent momentum classes. The system of equations can be solved for the form factors $G_1$ and $G_2$ by applying a two-parameter fit corresponding to minimizing the $\chi^2$-function given by $\chi^2 = (\bm{R}-M\bm{G})^T\ C^{-1}\ (\bm{R}-M\bm{G})$, where $C$ denotes the covariance matrix. Identifying $J_\mu$ with the vector current $V_\mu$ in \Eq{eq:ratio} yields the electromagnetic form factors $G_E$ and $G_M$, whereas the axial current $A_\mu$ leads to the axial vector form factors $G_A$ and $G_P$. We obtain the tensor charge $g_T$ at $Q^2=0$ from the tensor current $T_{\mu\nu}$, where only one independent ratio contributes.

\section{Setup}
This work makes use of the CLS $N_f=2+1$ $\mathcal{O}(a)$-improved Wilson fermion ensembles \cite{Bruno} listed in Tab.~\ref{tab:ensembles}. The tree-level improved L\"uscher-Weisz gauge action is employed in the gauge sector. These ensembles obey open boundary conditions in time in order to prevent topological freezing \cite{Luscher}. The physical quark masses are approached along a trajectory where the trace of the quark mass matrix is kept (approximately) constant, $\tr\ M_q = \text{const}$. To set the scale, we rely on the gradient flow time $t_0$ determined in Ref. \cite{Bruno2}.

\begin{table}[h]
\begin{center}
\begin{tabular}{l|ccccccccc}
	&$\beta$	&$a$ [fm] &$N_s^3\times N_t$	&$m_\pi$[MeV]	&$m_K$[MeV]	&$N_{\text{cfg}}^{\text{dis}}$	&$N_{\text{cfg}}^{\text{con}}$\\
\hline
H102	&3.40	&0.08636	&$32^3\times 96$	&352	&438	&997	&997\\
H105	&3.40	&0.08636 &$32^3\times 96$	&278	&460	&1019	&1019\\
C101	&3.40	&0.08636	&$48^3\times 96$	&223	&472	&1000	&2000\\
\hline
N401	&3.46	&0.07634	&$48^3\times 128$	&289	&462	&701	&701\\
\hline
N203	&3.55 &0.06426	&$48^3\times 128$ &345	&441	&768	&1536\\
N200	&3.55 &0.06426	&$48^3\times 128$	&283	&463	&852	&852\\
D200	&3.55 &0.06426	&$64^3\times 128$	&200	&480	&234	&936\\
\hline
N302	&3.70	&0.04981	&$48^3\times 128$	&354	&458	&1177	&1177\\
\end{tabular}
\end{center}
\caption{The ensembles used for this work. The last two columns give the number of configurations for the disconnected and connected diagrams, respectively.}
\label{tab:ensembles}
\end{table}

For the disconnected three-point function it is straightforward to check that it factorizes into separate traces of the quark loop and the two-point function
\begin{equation}
C_{3,J_\mu}^{N,l/s}(\bm{q},z_0;\bm{p}^\prime,y_0;\Gamma_\nu) = \left\langle \mathcal{L}_{J_\mu}^{l/s}(\bm{q},z_0)\cdot \mathcal{C}_2^N(\bm{p}^\prime,y_0;\Gamma_\nu) \right\rangle_G\ .
\end{equation}

The quark loops have been calculated with an improved stochastical estimator provided by hierarchical probing \cite{Stathopoulos}. Here the sequence of noise vectors is replaced by a sequence of Hadamard vectors, which are element-wise multiplied by one noise vector. In total, we considered two noise vectors, each multiplied with a sequence of 512 Hadamard vectors on every configuration and for each quark flavor. For the two-point function we used the standard nucleon interpolator 
\begin{equation}
N_{\alpha}(x) = \epsilon_{abc}\left( u_\beta^a(x)\ \left(C\gamma_5\right)_{\beta\gamma}\ d_\gamma^b(x) \right)\ u_\alpha^c(x)\ .
\end{equation}

To increase the overlap with the ground state, Wuppertal smearing \cite{Gusken} was used at the source and the sink. The statistical precision of the two-point function has been improved by employing the truncated-solver method \cite{Bali,Shintani}. For each ensemble we performed 32 low-precision solves on seven timeslices equally distributed around the center of the lattice and separated by seven timeslices without sources. On each of these timeslices one low-precision solve was considered for the bias correction, except for ensemble H105, where we considered four low-precision solves on each timeslice. For each current we considered all components combined with four choices of the projector $\Gamma$: $\Gamma_0 = \frac{1}{2}(1+\gamma_0)\ ,\ \Gamma_k = \frac{1}{2}(1+\gamma_0)\ \I\gamma_5\gamma_k\ ,\ k\in\{1,2,3\}$. Details on the two- and three-point functions for the connected contributions at $Q^2=0$ can be found in Ref. \cite{Harris}. The same details also carry over to $Q^2\neq 0$. Here, we did not make use of the zeroth component of the axial vector current, due to its large excited-state contamination. Furthermore, the electromagnetic form factors can be obtained directly, as shown in Ref. \cite{Capitani}, without fitting the system in \Eq{eq:system}. Note that for the connected three-point function the nucleon at the sink is always at rest, whereas in the disconnected contributions we consider up to two units of the squared lattice integer momentum. For the local axial vector and the conserved vector current we implemented the improved versions
\begin{equation}
A_\mu(\vec{z},z_0)^{\text{Imp.}}=A_\mu(\vec{z},z_0)+ac_A\ \partial_\mu P(\vec{z},z_0)\ ,\ V_\mu(\vec{z},z_0)^{\text{Imp.}}=V_\mu(\vec{z},z_0)+ac_V\ \partial_\nu T_{\nu\mu}(\vec{z},z_0)\ ,
\end{equation}
where the improvement coefficients have been determined non-perturbatively in Ref. \cite{Bulava,Gerardin}. For the tensor charge we used the local tensor current.

\section{Renormalization}
As starting point for the renormalization procedure we consider the flavor-diagonal operators 
\begin{equation}
O^a_\Gamma(x) = \bar{\psi}(x) \Gamma \lambda^a \psi(x)\ \ ,\ \ \psi = (u,d,s)^T\ \ ,\ \ a\in\{3,8,0\}\ ,
\end{equation}
\begin{equation}
\lambda^3 = \dfrac{1}{\sqrt{2}}\text{diag}\left(1,-1,0\right)\ ,\ \lambda^8 = \dfrac{1}{\sqrt{6}}\text{diag}\left(1,1,-2\right)\ ,\ \lambda^0 = \dfrac{1}{\sqrt{3}}\text{diag}\left(1,1,1\right)\ .
\end{equation}

We make use of $N_f=3$ $\mathcal{O}(a)$-improved Wilson fermion ensembles, and hence, the renormalization matrix of the flavor-diagonal operators is given by
\begin{equation}
Z_\Gamma = \text{diag}\left(Z_\Gamma^{33},Z_\Gamma^{88},Z_\Gamma^{00}\right)\ \ ,\ \ Z_\Gamma^{33} = Z_\Gamma^{88}\ \ .
\end{equation}

All details on our renormalization procedure for the $Z_\Gamma^{33}$ can be found in Ref. \cite{Harris}. We follow the same procedure for $Z_\Gamma^{00}$ which, however, necessitates the calculation of additional diagrams. The required quark loops have been estimated with hierarchical probing, using 512 Hadamard vectors on each configuration. Furthermore, in the case of the singlet axial operator, an anomalous dimension arises (given in Ref. \cite{Larin}) and the conversion factors $Z^{\overline{\text{MS}}}_{\text{RI}^\prime\text{-MOM}}$ for the singlet vector and axial vector operators are only known to one loop. Finally, we apply a basis transformation to the physical basis given by
\begin{equation}
\left(\begin{array}{c}
O_\Gamma^{u-d}(x)_R\\
O_\Gamma^{u+d}(x)_R\\
O_\Gamma^{s}(x)_R\\
\end{array}  \right) = \left(\begin{array}{ccc} Z_\Gamma^{u-d,u-d} &0 &0\\ 0 &Z_\Gamma^{u+d,u+d} &Z_\Gamma^{u+d,s}\\ 0 &Z_\Gamma^{s,u+d} &Z_\Gamma^{s,s} \end{array}\right)\ \left(\begin{array}{c}
O_\Gamma^{u-d}(x)\\
O_\Gamma^{u+d}(x)\\
O_\Gamma^{s}(x)\\
\end{array}  \right)\ .
\end{equation}

The final renormalization constants are taken in the $\overline{\text{MS}}$-scheme at $\mu=2\,\text{GeV}$. Note that the renormalization constants at $\beta=3.7$ have been obtained from a linear extrapolation. To account for the systematic uncertainty introduced through the extrapolation, the error of the renormalization constants at this $\beta$ has been inflated by a factor of ten. The reliability of the extrapolation was checked for the isovector axial charge $g_A$ in Ref. \cite{Harris}, where consistent results from our renormalization procedure compared to renormalizing through the Schr\"odinger functional were found.

\section{Results}
\begin{figure}
\hspace{1cm}
\includegraphics[scale=0.39]{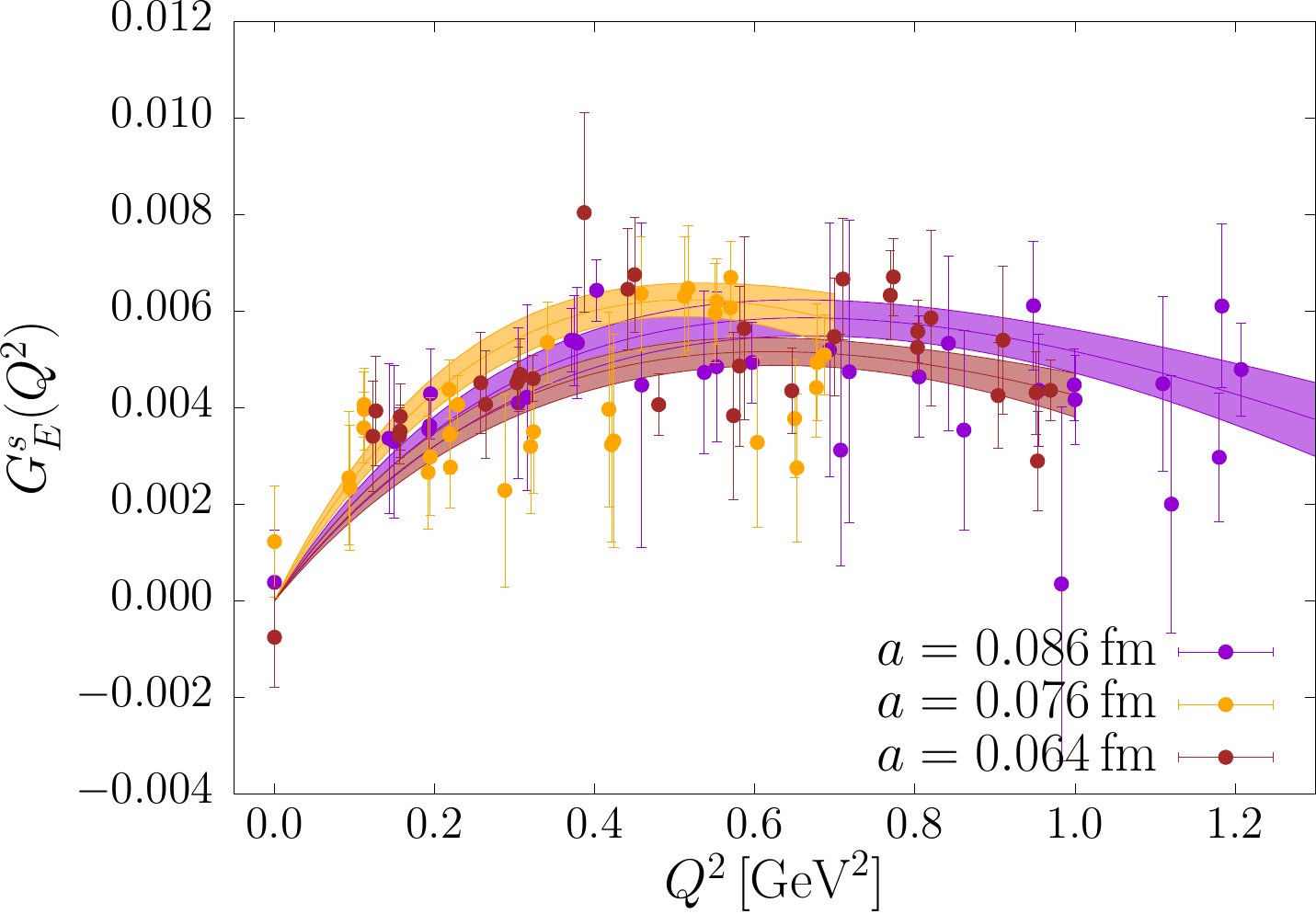}
\hspace{1cm}
\includegraphics[scale=0.39]{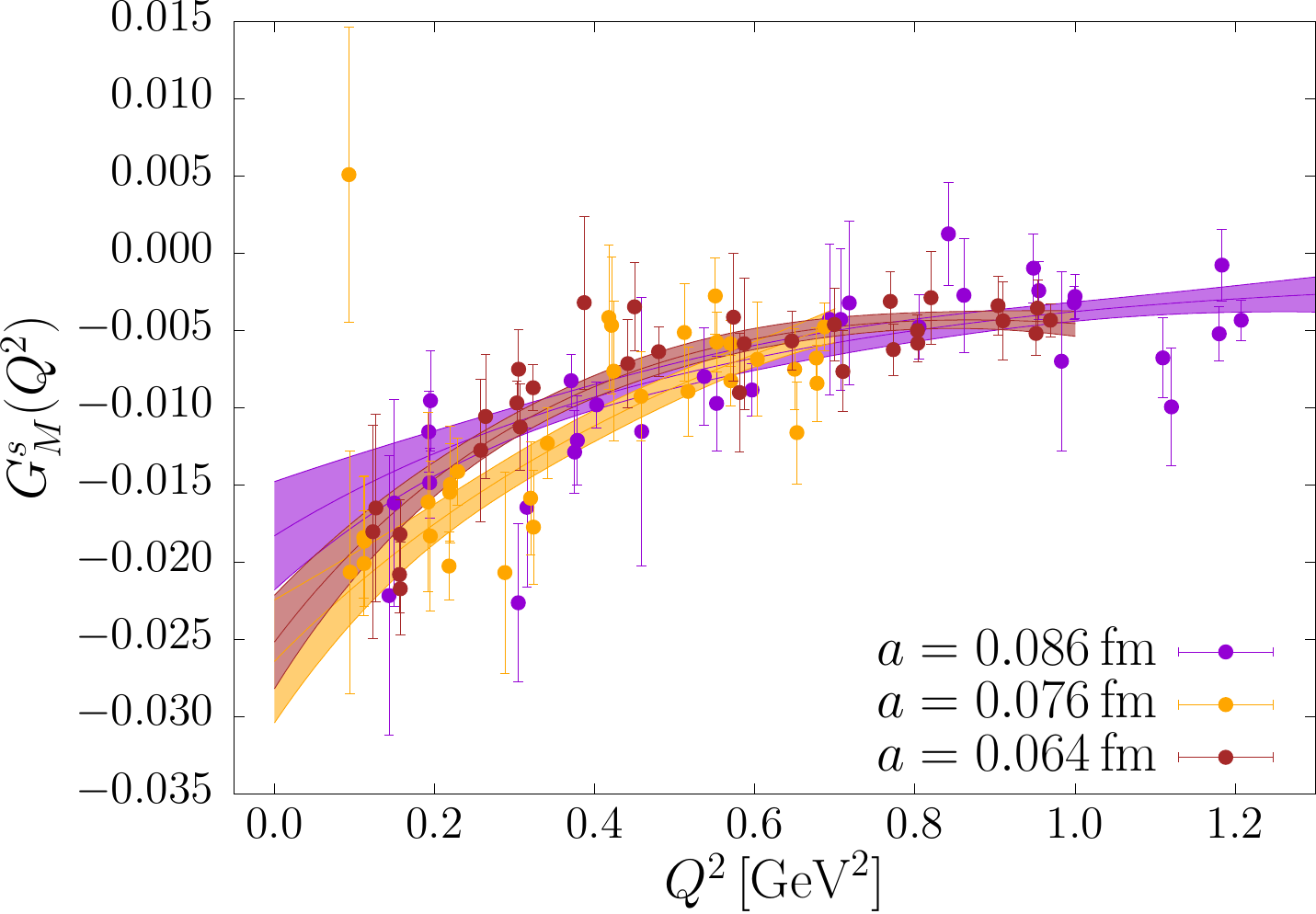}
\caption{The $Q^2$-dependence of the strange electromagnetic form factors at a kaon mass of $m_K\approx 460\,\text{MeV}$.}
\label{fig:svecff}
\end{figure}

In \Fig{fig:svecff}, we show the resulting strange electromagnetic form factors for three lattice spacings at constant kaon mass of $m_K\approx 460\,\text{MeV}$. Excited-state contamination has been handled with the summation method with $y_0\in[0.5,1.3]\,\text{fm}$. The bands correspond to $z$-expansion fits \cite{Hill} to fifth order using the two-kaon production threshold for the conformal map, where Gaussian priors of the form $\tilde{a}_k = 0 \pm 5\max\{|a_0|,|a_1|\}\ \forall\ k>1$ have been employed to stabilize the fits. 

\begin{figure}
\includegraphics[scale=0.325]{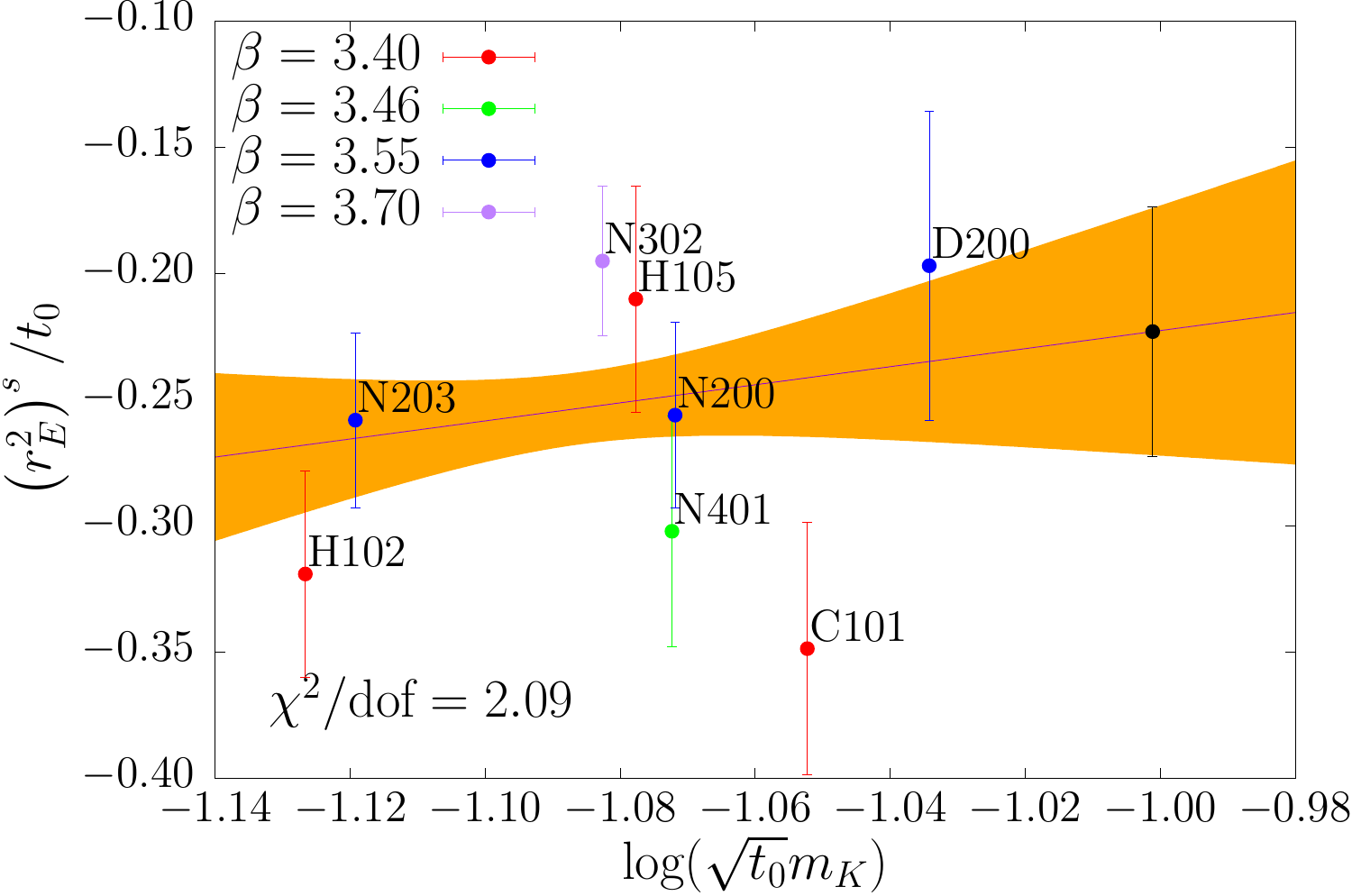}
\includegraphics[scale=0.325]{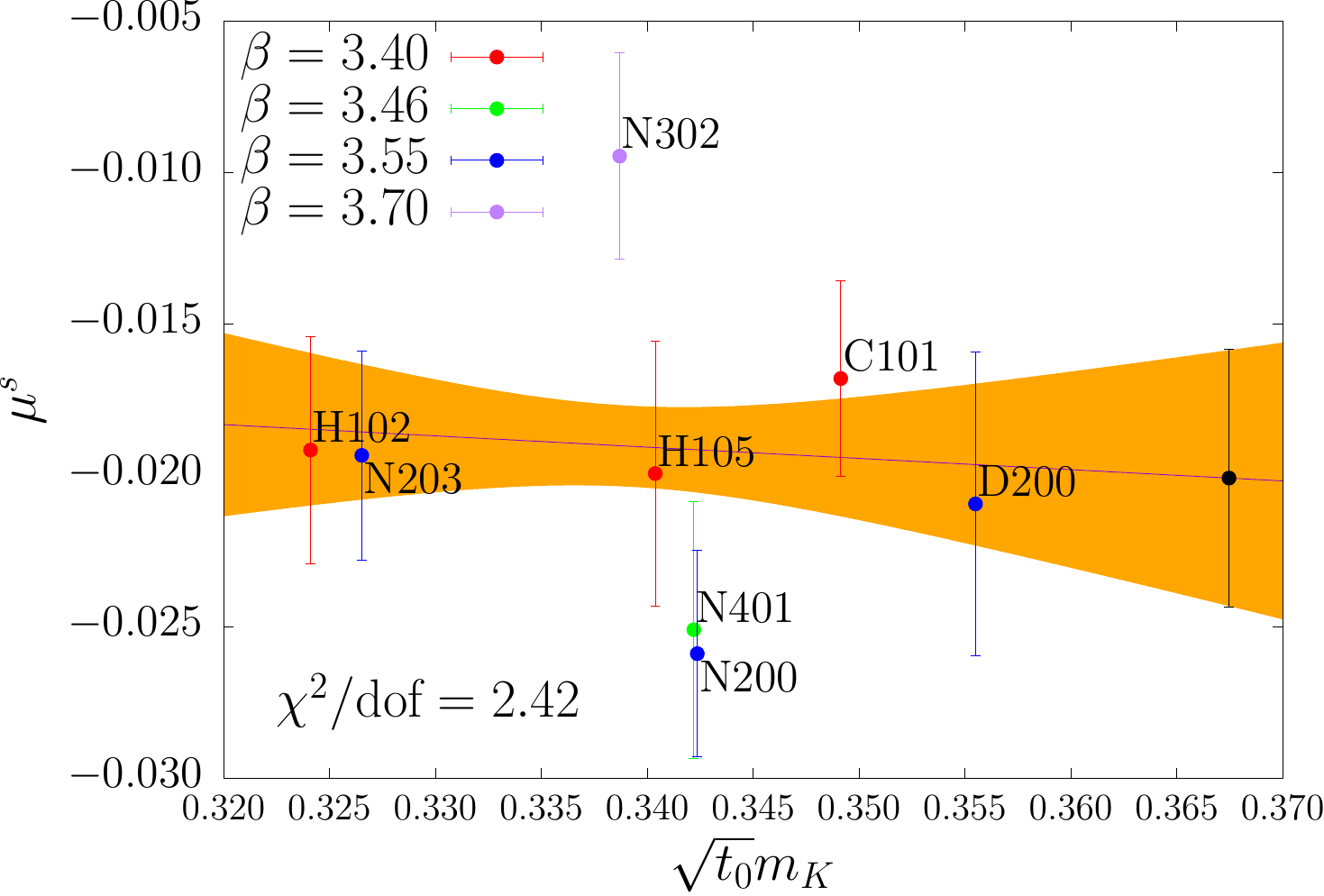}
\includegraphics[scale=0.325]{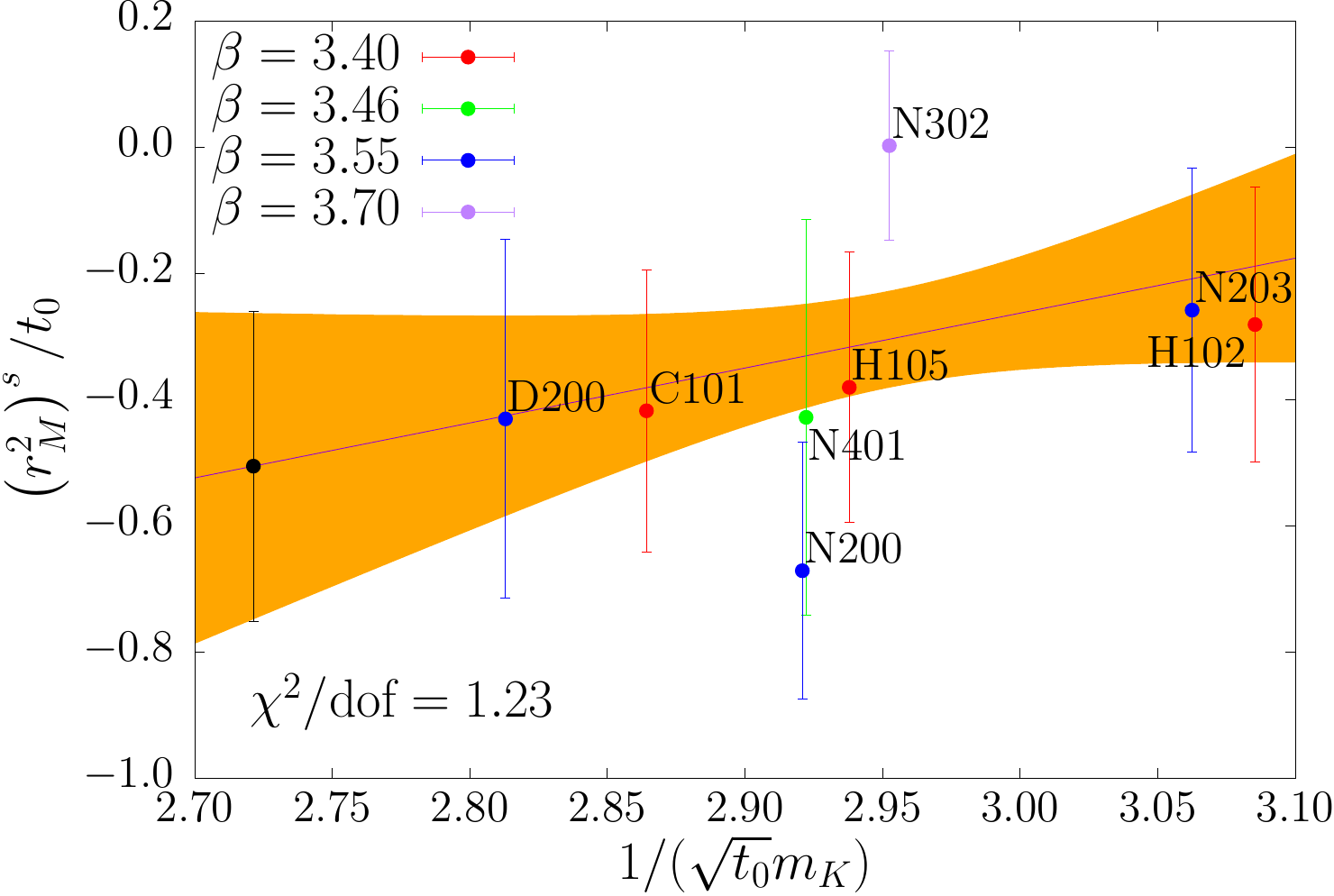}
\caption{The kaon mass dependence of the strange magnetic moment and strange charge radii. The orange band is a linear extrapolation to the physical kaon mass (black data point).}
\label{fig:extsEMff}
\end{figure}

From the $z$-expansion fits we can extract the strange magnetic moment $\mu^s=G^s_M(0)$ and the strange electromagnetic charge radii $(r^2)_{E/M}^s$, shown in \Fig{fig:extsEMff}. Note that for our main result we impose a cut in $Q^2$ at $0.5\,\text{GeV}^2$. We extrapolate to the physical kaon mass by performing linear fits of the form 
\begin{equation}
(r^2_E)^s(m_K) = a_0 + a_1 \log(m_K)\ ,\ \mu^s(m_K) = a_2 + a_3 m_K\ ,\ (r^2_M)^s(m_K) = a_4 + a_5/m_K\ ,
\end{equation}
derived from leading-order SU(3) HBChPT \cite{Hemmert}, where we add the higher-order $a_4$-term as our data does not show the leading-order divergence expected by HBChPT. 

In order to estimate systematic errors we consider four variations: the inclusion of $\mathcal{O}(a^2)$ lattice artifacts, doubling the prior width in the $z$-expansion fits, handling excited states with the plateau method at $\sim 1\,\text{fm}$, and removing the cut in $Q^2$. The respective errors are taken from the difference to the main result and then added in quadrature leading to our final result
\begin{equation}
(r^2_E)^s = -0.0048(11)(23)\,\text{fm}^2\ ,\ \mu^s = -0.020(4)(11)\ ,\ (r^2_M)^s = -0.011(5)(12)\,\text{fm}^2\ .
\end{equation}

\begin{figure}
\includegraphics[scale=0.325]{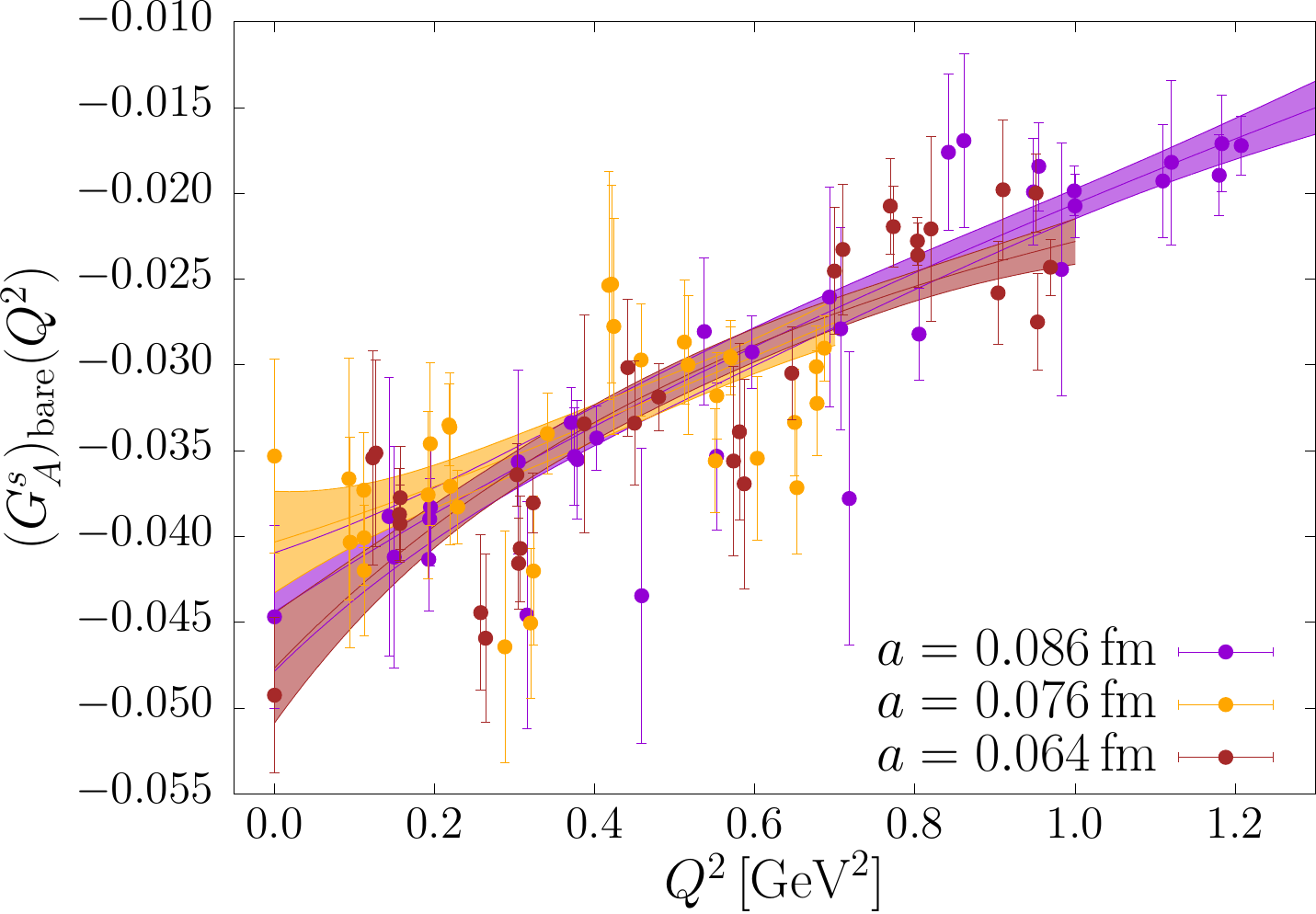}
\includegraphics[scale=0.325]{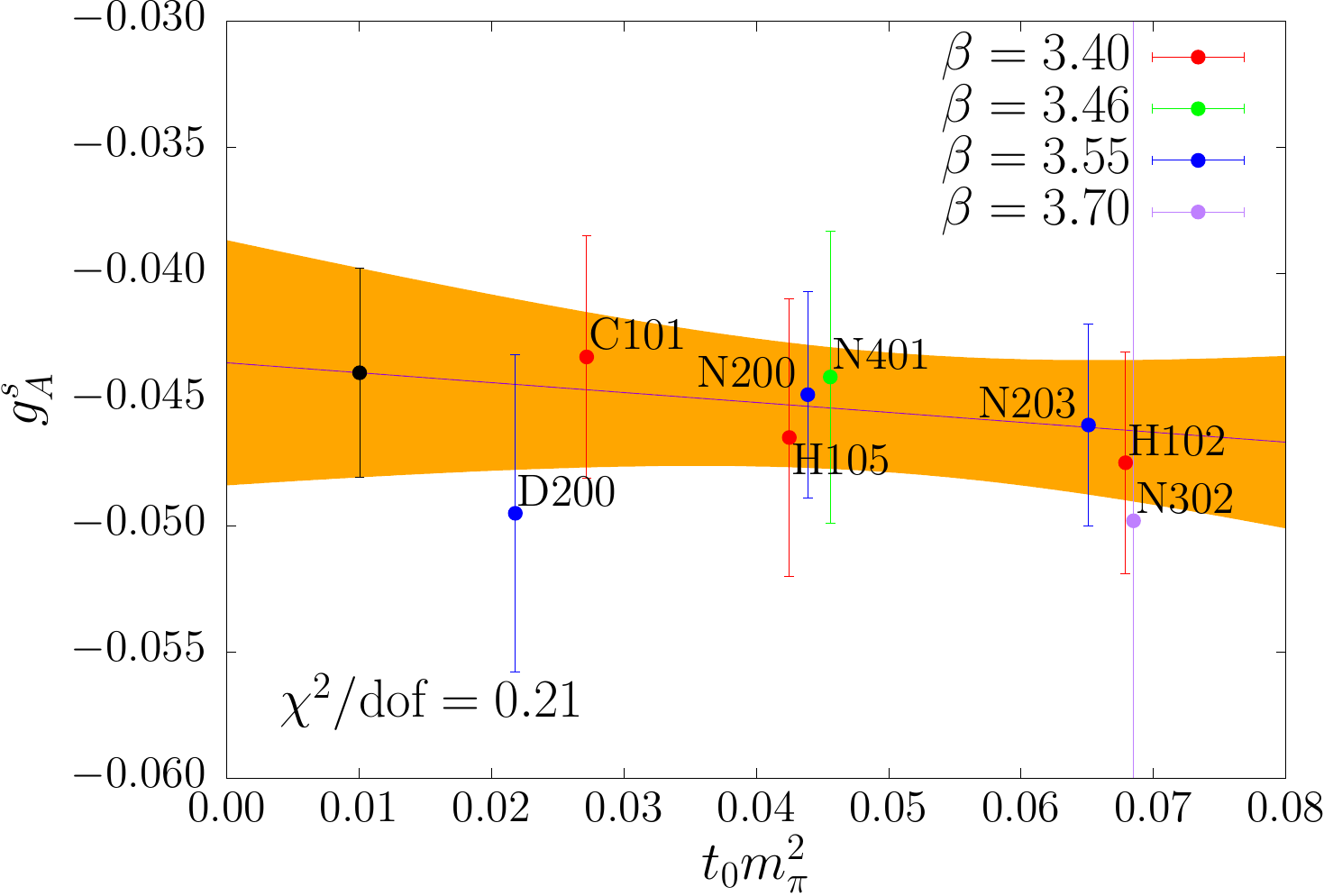}
\includegraphics[scale=0.325]{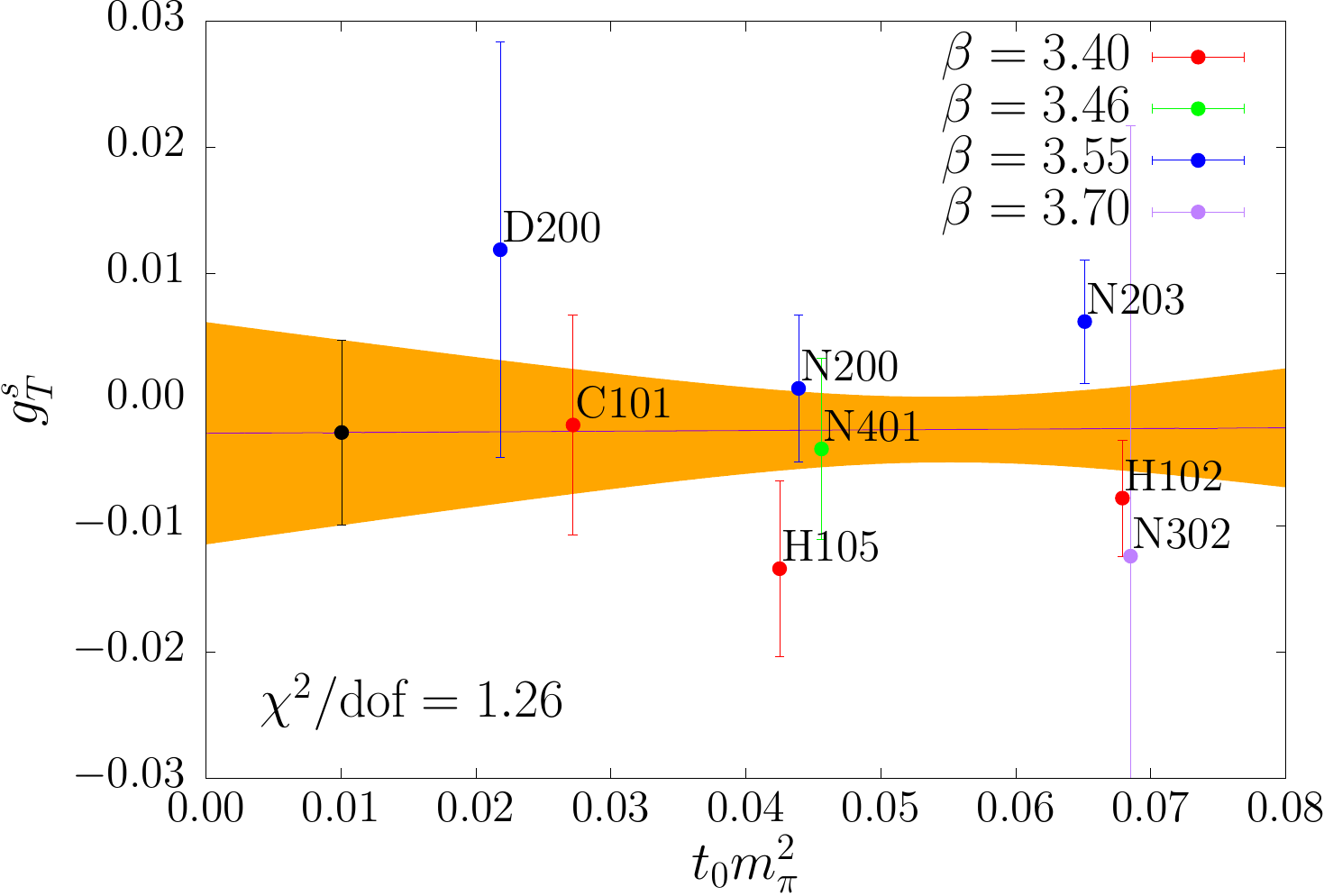}
\caption{The $Q^2$-dependence of the strange axial vector form factor for three lattice spacings at constant kaon mass of $m_K\approx 460\,\text{MeV}$ (left), and the extrapolation of the strange axial charge (middle) and strange tensor charge (right) to the physical point (black data point).}
\label{fig:saxvecffagts}
\end{figure}

We follow the same procedure for the strange axial vector form factor, shown in \Fig{fig:saxvecffagts}. Also shown is a linear extrapolation in $m_\pi^2$ to the physical point for the strange contribution to the axial charge. The same sources of systematics are considered in this case with the adjustment to $\mathcal{O}(a)$ lattice artifacts, as the renormalization constants are not $\mathcal{O}(a)$-improved. Our final result is given by $g_A^s = -0.044(4)(5)$. Furthermore, in \Fig{fig:saxvecffagts}, we show the analogous linear extrapolation of the strange tensor charge, again obtained from the summation method. In the estimation of the systematic error, we consider the plateau method at $\sim 1\,\text{fm}$ and the inclusion of $\mathcal{O}(a)$ lattice artifacts in the extrapolation. The final result is $g_T^s = -0.0026(73)(424)$.

We make use of the following decomposition of the $u$- and $d$-charges
\begin{equation}
g^u_{A/T} = \frac{1}{2}\left( g_{A/T}^{u+d-2s} + 2 g_{A/T}^{s} + g_{A/T}^{u-d} \right)\ ,\ g^d_{A/T} = \frac{1}{2}\left( g_{A/T}^{u+d-2s} + 2 g_{A/T}^{s} - g_{A/T}^{u-d} \right)\ .
\label{eq:qcdec}
\end{equation}

\begin{figure}
\includegraphics[scale=0.325]{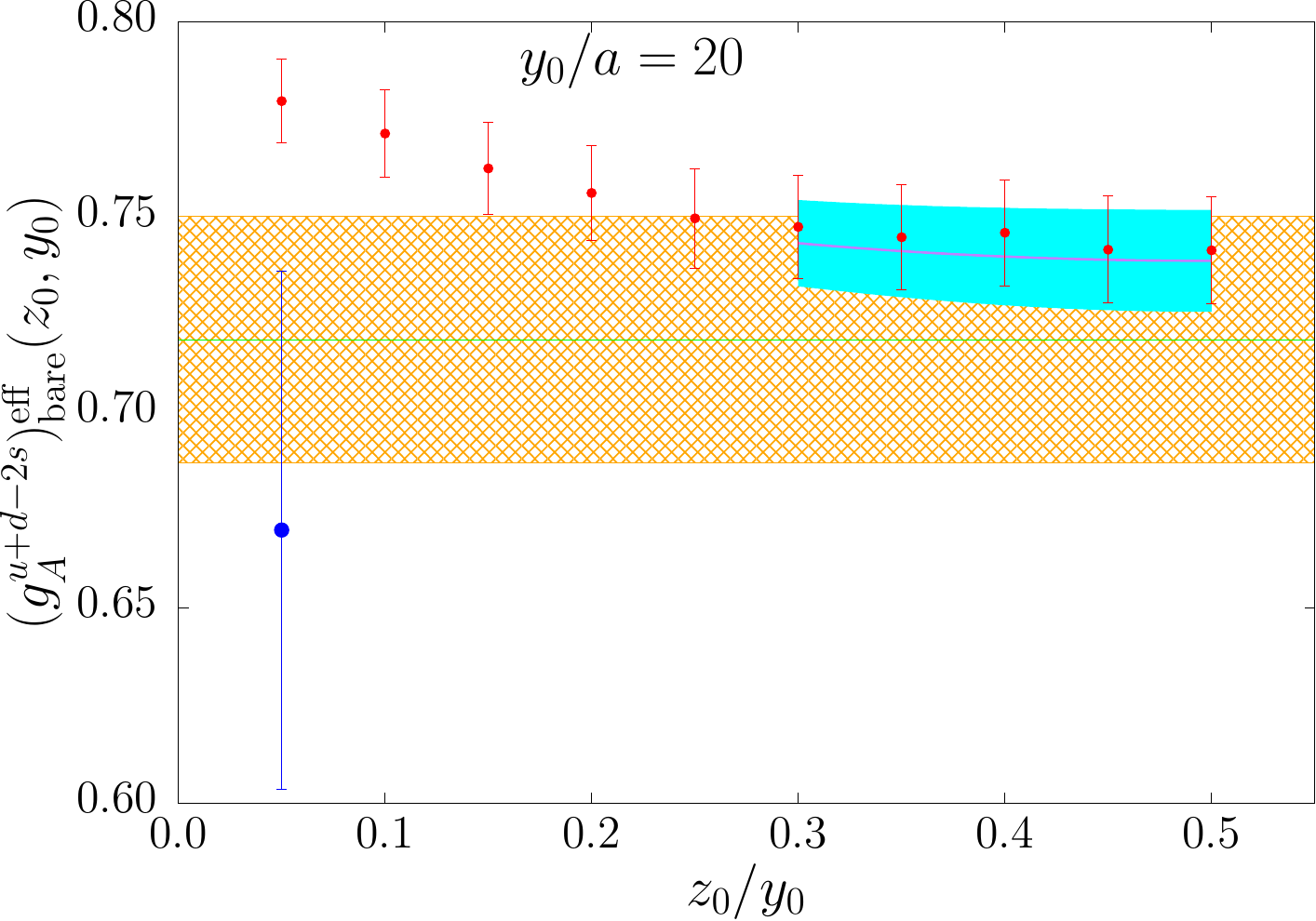}
\includegraphics[scale=0.325]{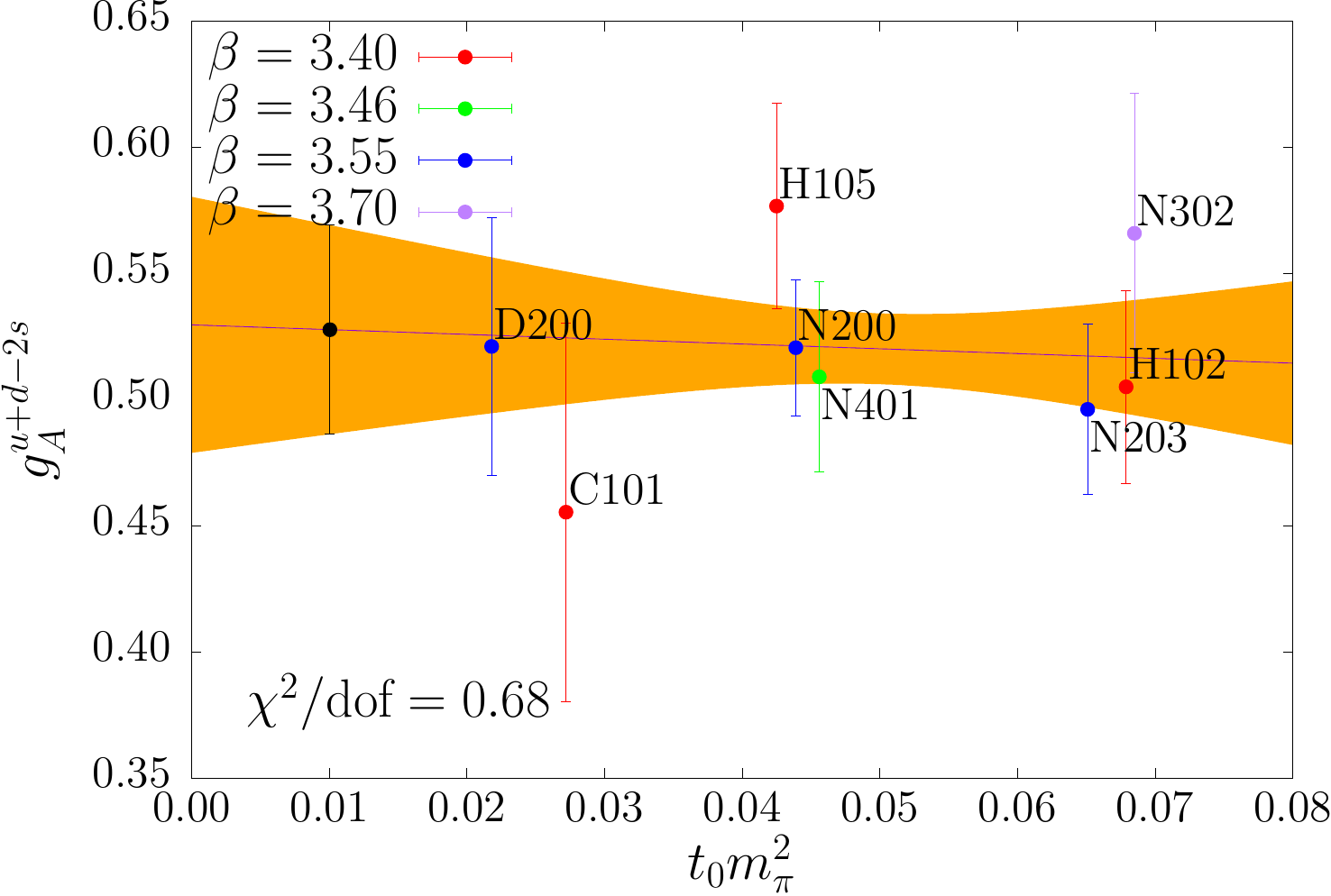}
\includegraphics[scale=0.325]{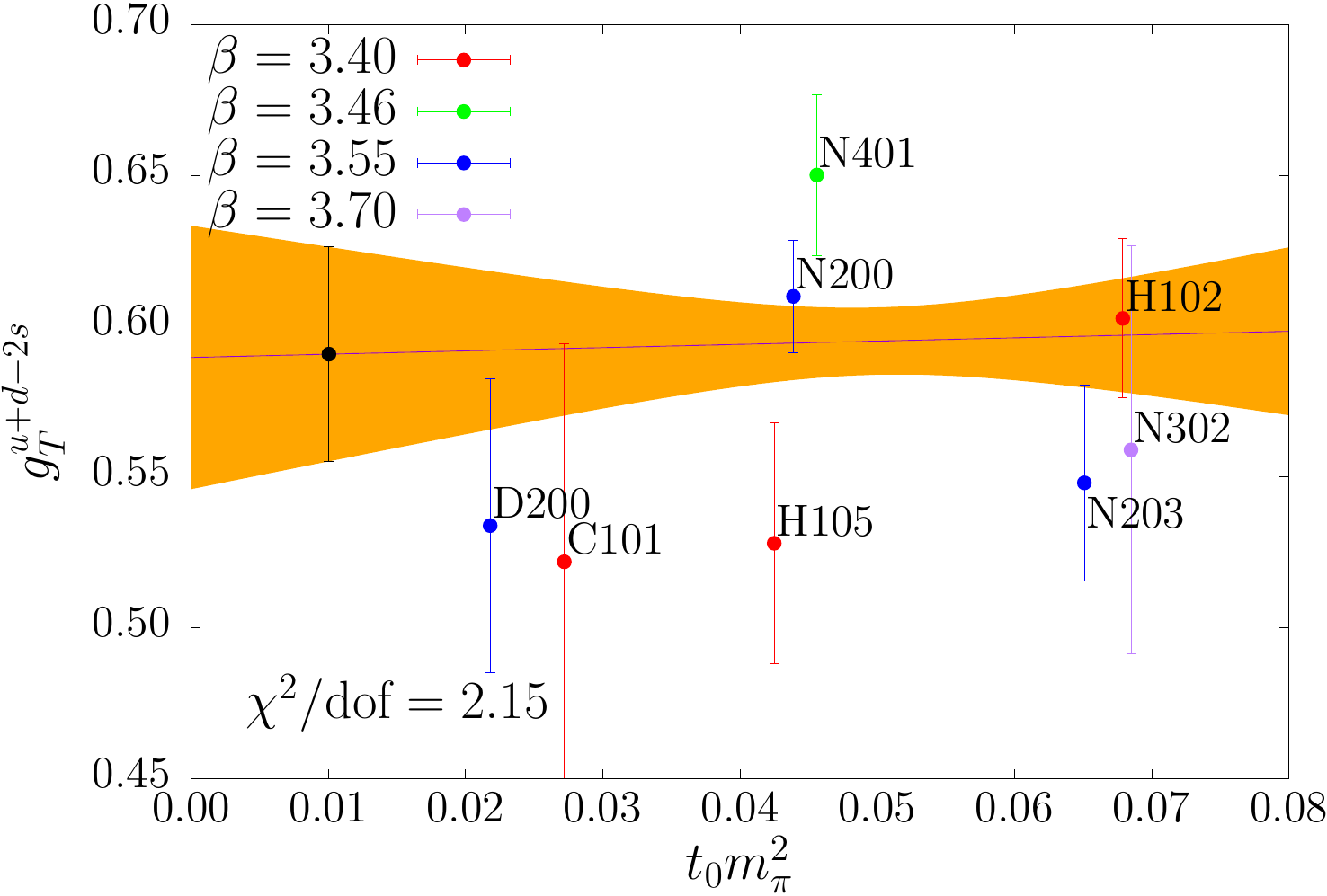}
\caption{Two-state fit to $g_A^{u+d-2s}$ at the smallest $y_0$ on ensemble N302 (left), where also the summation method result is shown (blue data point), and extrapolation of $g_A^{u+d-2s}$ (middle) and $g_T^{u+d-2s}$ (right) to the physical point (black data point).}
\label{fig:qcat}
\end{figure}

The strange contributions have been shown above and our results for the isovector charges can be found in Ref. \cite{Harris}. It is favorable to consider the $(u+d-2s)$-contribution, as it renormalizes like the isovector charges, so that no mixing occurs, and the disconnected contributions can be combined to $(l-s)$, which leads to a noise reduction. To handle the excited state contamination, we perform two-state fits simultaneously to all source-sink separations $y_0$ of the form
\begin{equation*}
g_{A/T}(z_0,y_0) = g_{A/T} + A_{A/T} \left( e^{-\Delta z_0} + e^{-\Delta (y_0 - z_0)} \right) + B_{A/T} e^{-\Delta y_0}\ ,
\end{equation*}
where the energy gap $\Delta$ to the first excited state has been determined in a simultaneous two-state fit to all six isovector charges, as explained in detail in Ref. \cite{Harris}. We show an example of this fit in \Fig{fig:qcat}, where we also compare to the summation method. We then perform linear extrapolations in $m_\pi^2$ to the physical point, also shown in \Fig{fig:qcat}.

To assign a systematic error, we consider the difference to the extrapolation of the summation method results and an extrapolation with the inclusion of $\mathcal{O}(a)$ lattice artifacts. The final results are given by
\begin{equation}
g_A^{u} = 0.84(3)(4)\ ,\ g_A^{d} = -0.40(3)(4)\ ,\ g_T^{u} = 0.77(4)(6)\ ,\ g_T^{d} = -0.19(4)(6)\ .
\end{equation}

\paragraph{Acknowledgements:} This research is supported by the DFG through the SFB 1044, and under grant HI 2048/1-1. Calculations for this project were partly performed on the HPC clusters "{}Clover"{} and "{}HIMster II"{} at the Helmholtz-Institut Mainz and "{}Mogon II"{} at JGU Mainz. Additional computer time has been allocated through projects HMZ21 and HMZ36 on the BlueGene supercomputer system "{}JUQUEEN"{} at NIC, J\"ulich. Our programmes use the QDP++ library \cite{QDPpp} and deflated SAP+GCR solver from the openQCD package \cite{openQCD}, while the contractions have been explicitly checked using \cite{QCT}. We are grateful to our colleagues in the CLS initiative for sharing ensembles. The ensembles for the calculation of the renormalization constants have been generated in a joint effort with the RQCD collaboration.

\end{document}